\documentclass[aps,prd,twocolumn,groupedaddress,amsmath,amssymb,floatfix]{revtex4-2}
\usepackage{lineno,graphicx,color,hyperref,bm, orcidlink,placeins,todonotes}

\begin{document}
\preprint{APS/123-QED}

\title{Benchmarks and applications of the nuclear deexcitation event generator \textsc{NucDeEx}}

\newcommand{\kamioka}{\affiliation{Kamioka Observatory, Institute for Cosmic Ray Research, \\
University of Tokyo, Kamioka, Gifu 506-1205, Japan}}
\author{Seisho Abe\,\orcidlink{0000-0002-2110-5130}} 
\email{seisho@hep.phys.s.u-tokyo.ac.jp}
\kamioka
\thanks{Present address: Department of Physics, Graduate School of Science, The University of Tokyo, 7-3-1 Hongo, Bunkyo-ku, Tokyo, 113-0033, Japan}
\date{\today}

\begin{abstract}
Neutron multiplicity is a key observable in recent neutrino experiments that can enhance the sensitivity of various neutrino physics searches.
Nuclear deexcitation plays a significant role in neutron emissions associated with neutrino-nucleus interactions. 
Therefore, precise prediction of this process is essential.
To address this need, a general-purpose nuclear deexcitation event generator \textsc{NucDeEx} was developed and released as an open-source package.
The treatment of low-lying discrete excited states was updated to better reproduce experimental data.
Benchmarks were conducted using existing nuclear deexcitation event generators and experimental data.
Application to other simulators, neutrino event generator \textsc{NEUT} and general particle simulation tool \textsc{Geant4}, are also presented.
\end{abstract}
\maketitle

\section{Introduction} \label{sec:introduction}
Neutron multiplicity has emerged as a key observable in recent neutrino detectors.
Large liquid scintillator detectors such as KamLAND~\cite{PhysRevD.107.072006} and JUNO~\cite{An_2016} are sensitive to gamma rays from nuclear deexcitation and thermal neutron capture because of their high light yield.
In water Cherenkov detectors such as Super-Kamiokande (SK), the detection efficiency of the 2.2\,MeV gamma ray from thermal neutron capture on hydrogen is limited to about 20\% due to its small light yield~\cite{Abe_2022_SK}.
Super-Kamiokande Gadolinium (SK-Gd)~\cite{ABE2022166248,ABE2024169480} successfully enhanced the detection efficiency for the thermal neutron capture by dissolving Gd in water, which emits gamma rays with a total energy of approximately 8\,MeV.
Precise prediction of neutron multiplicity associated with neutrino-nucleus interactions is particularly important in searches for the diffuse supernova neutrino background (DSNB)~\cite{PhysRevD.104.122002,Harada_2023,Abe_2022}.
In their analyses, the signal is inverse beta decay ($\bar{\nu}_e+p\rightarrow e^+ + n$) whereas the dominant background arises from atmospheric neutrino interactions.
\par
Nuclear deexcitation plays a significant role in neutron emissions associated with neutrino-nucleus interactions.
KamLAND reported good agreement between the predicted and observed neutron multiplicities associated with atmospheric neutrino interactions~\cite{PhysRevD.107.072006}.
Notably, they coupled their custom deexcitation event generator with the neutrino event generator \textsc{NuWro}~\cite{PhysRevC.86.015505} in this study.
The T2K far detector (SK) analysis reported significant overprediction in neutron multiplicity of charged-current quasielastic (CCQE)-enriched sample~\cite{Akutsu:2019fsn}.
However, their recent investigation on NCQE interactions observed that the overprediction is largely mitigated by using alternative secondary interaction models in detector simulations~\cite{qh28-4znk}.
SK-Gd also published several results recently on neutron multiplicity measurements associated with atmospheric neutrino interactions, showing that the observed multiplicities fall within the wide range predicted by various models, particularly nuclear deexcitation models~\cite{PhysRevD.109.L011101,4d71-d69k}.
The reason for overprediction of neutron multiplicity at SK(-Gd) has been gradually identified by these studies.
The deexcitation model coupled with the Bertini cascade model~\cite{WRIGHT2015175}, which had long been used in SK detector simulation, was overly simple and lacked sufficient descriptive power.
Not limited to these experiments, SNO~\cite{PhysRevD.99.112007} and MINERvA~\cite{PhysRevD.100.052002,PhysRevD.108.112010} measured neutron multiplicity, and extensive investigation on model variations in large liquid scintillator detectors was conducted mainly for JUNO~\cite{PhysRevD.103.053001, PhysRevD.103.053002,Cheng2025}.
Beyond neutrino interaction, nuclear deexcitation modeling is also essential for predicting nucleon decay in nuclei~\cite{PhysRevC.48.1442,PhysRevD.67.076007,Hagino_2018}.
Since deexcitation emits not only neutrons but also gamma rays, it has a significant impact on the detector response in experiments such as SK~\cite{PhysREvD.100.112009}.
\par
The reduction of systematic uncertainties from neutrino-nucleus interactions is becoming increasingly critical in next-generation experiments such as Hyper-Kamiokande~\cite{Hyper-Kamiokande:2025fci} and DUNE~\cite{PhysRevD.105.072006}.
These experiments require more accurate modeling of final-state particles and nuclear deexcitation products to reduce systematic uncertainties.
Recent measurements of low-energy $\alpha$ and $\gamma$ particles in the ArgoNeuT and MicroBooNE liquid-argon time projection chambers~\cite{PhysRevD.99.012002,PhysRevD.109.052007} indicate that such detectors can measure particles originating from nuclear deexcitation.
Such measurements provide powerful benchmarks for models describing nuclear deexcitation.
\par
To maximize the performance of these neutrino detectors, a reliable and general-purpose nuclear deexcitation event generator that can be used in combination with neutrino event generators is essential.
Several attempts have been made to incorporate nuclear deexcitation processes within the framework of neutrino event generators:
for example, the use of ABLA~\cite{PhysRevC.105.014623,PhysRevC.98.021602,BENLLIURE1998458,JUNGHANS1998635} within the INCL++ (the Li\'{e}ge intranuclear cascade model)~\cite{PhysRevC.87.014606,Leray_2013} in \textsc{NuWro}~\cite{PhysRevD.108.112008}, the PEANUT model in \textsc{FLUKA}~\cite{BATTISTONI201510}, and low-energy neutrino event generator \textsc{MARLEY}~\cite{GARDINER2021108123}.
However, no tool has yet become widely adopted in neutrino physics analyses.
This highlights the need for a flexible and well-validated deexcitation event generator that can be readily integrated into simulations.
\par
In response to this need, a general-purpose nuclear deexcitation event generator \textsc{NucDeEx} was developed and released as an open-source package~\cite{PhysRevD.109.036009,code}.
Following the initial release, several updates on its algorithm and more detailed benchmarks against existing nuclear deexcitation event generators and experimental data were conducted.
Furthermore, \textsc{NucDeEx} has recently been integrated into the neutrino event generator \textsc{NEUT}~\cite{Hayato2021}, while user-side implementations have been made for \textsc{Geant4}~\cite{AGOSTINELLI2003250,code_g4}, enabling more precise simulations of nuclear deexcitation in neutrino-nucleus interactions and detector simulations.
This work addresses the long-standing need for validated modeling of nuclear deexcitation in neutrino detectors, aiming to improve predictive power for various physics purposes.
This paper presents recent developments of \textsc{NucDeEx} (Sec.~\ref{sec:updates}), comprehensive comparisons with other deexcitation event generators (Sec.~\ref{sec:validation}), and its applications to other simulations (Sec.~\ref{sec:applications}).

\section{Highlights of updates} \label{sec:updates}
The latest version \textsc{NucDeEx} 2.2 was released on June 2025.
This section summarizes major updates.

\subsection{Discrete excited states} \label{sec:phole}
\begin{figure*}[htbp]\centering
\includegraphics[width=0.75\textwidth]{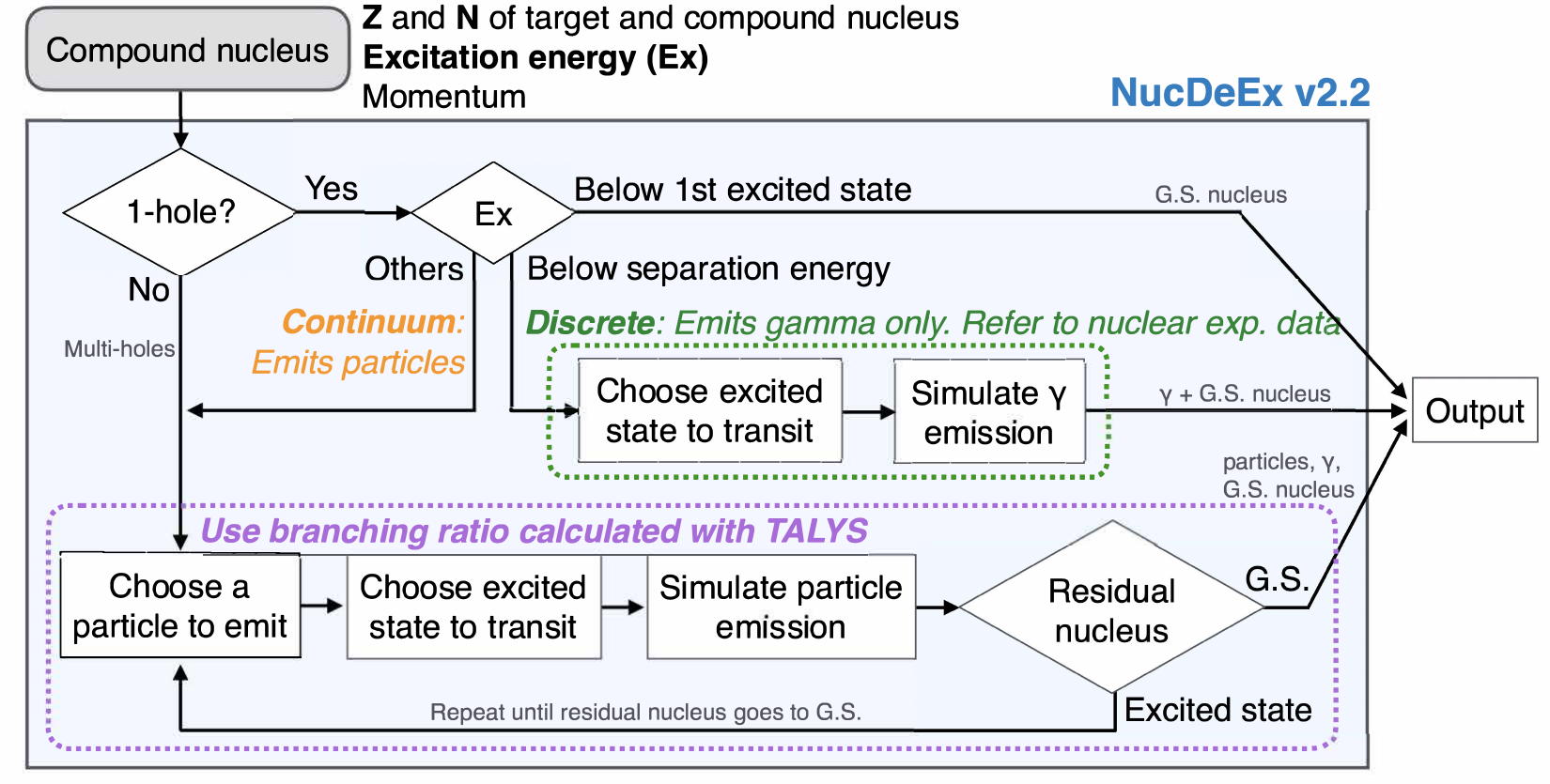}
\caption{Algorithm of \textsc{NucDeEx} version 2.
         The excited state to which the transit occurs is determined according to the branching ratios in Table~\ref{tab:br_p_16O}.
         This method is designed to be used with the SF~\cite{BENHAR1994493,PhysRevD.72.053005} or INCL++~\cite{PhysRevC.87.014606,Leray_2013}, which do not have clear peaks corresponding to low-lying discrete excited states.
         }
\label{fig:algorithm}
\end{figure*}

\textsc{NucDeEx} was originally designed to be used with the spectral function (SF)~\cite{BENHAR1994493,PhysRevD.72.053005}.
Since the SF does not have the resolution to separate low-lying discrete excited states, the excited state (hole state) must be determined within the \textsc{NucDeEx} framework for this region.
In version 1, the hole states of $p_{1/2},\,p_{3/2},\,$ and $s_{1/2}$ were determined by applying a box-cut condition on the excitation energy (Table~IV in Ref.~\cite{PhysRevD.109.036009}).
This cut overpredicted the probabilities of the hole state of $p_{1/2}$ and $p_{3/2}$, because it treated all excitation energies below 16\,MeV as either of the two hole states.
To address this issue, version 2 introduces a new scheme for classifying excitation energies (Fig.~\ref{fig:algorithm}):
\begin{enumerate}
  \item Energies below the first excited state are treated as the ground state.
  \item Energies between the first excited state and the separation energy are treated as discrete excited states.
  \item Energies above the separation energy are considered as continuum excited states.
\end{enumerate}
Decays from the continuum excited states are described by the Hauser-Feshbach model~\cite{PhysRev.87.366}, as implemented in the nuclear reaction code \textsc{TALYS}~\cite{Koning2023}, while those from discrete excited states are generated according to the branching ratios described below.

The gamma-ray branching ratios from the discrete excited states of $^{15}\text{N}^*$ and $^{15}\text{O}^*$ have also been updated (Table~\ref{tab:br_p_16O}).
For example, in the case of $^{15}\text{N}^*$, version 1 considered two excited states at 6.32\,MeV and 9.93\,MeV below the proton separation energy based on Ref.~\cite{PhysRevC.48.1442}.
Experimental data, however, indicate a non-negligible contribution from the excited states at 5.27\,MeV and 5.30\,MeV~\cite{YosM:2003, PhysRevC.49.955}.
Thus, version 2 considers a 9\% contribution from these excited states, treating them as a single level due to their small energy separation.
The minor contributions from excited states at 8.31\,MeV and 9.05\,MeV, although observed in Ref.~\cite{YosM:2003}, are neglected here because of their minimal impact.
The excited state at 10.70\,MeV is treated as part of the continuum excited states in version 2.
Due to the absence of experimental data, the branching ratios for $^{15}$O$^*$ are inferred from those of $^{15}$N$^*$, assuming that states with the same spin-parity and similar excitation energy exhibit analogous branching ratios.
The excited state at 9.61\,MeV of $^{15}\text{O}^*$, which corresponds to the 9.93\,MeV state of $^{15}\text{N}^*$, lies above the separation energy and is therefore not considered as a discrete excited state here.
The branching ratios for $^{11}\text{B}^*$ and $^{11}\text{C}^*$ remain unchanged from Table~I in Ref.~\cite{PhysRevD.109.036009}.

\begin{table*}[htbp] \centering
\caption{Excited states and branching ratios for discrete levels of $^{15}\text{N}^{*}$ and $^{15}\text{O}^{*}$ used in NucDeEx version 2.
         The excitation energy and gamma-ray energy are denoted as $E_x$ and $E_\gamma$, respectively.
         The branching ratios (Br) for $^{15}\text{N}^{*}$ are from~\cite{YosM:2003, PhysRevC.49.955}.
         The excited states of $^{15}\text{N}^*$ at 5.27\,MeV and 5.30\,MeV, as well as those of $^{15}\text{O}^*$ at 5.18\,MeV and 5.24\,MeV, are grouped together as a single level.
         Gamma-ray emission from these excited states may include multiple transitions.
         The relative branching ratios for gamma rays (RBr$_\gamma$) are based on \textsc{TALYS}~\cite{Koning2023}.
         Only those with RBr$_\gamma>0.01$ are listed.
         }
\label{tab:br_p_16O}
\begin{tabular*}{1.0\textwidth}{@{\extracolsep{\fill}}lccccc} \hline \hline
Nucleus & $E_x$\,(MeV) & $J^\pi$ & Br & RBr$_\gamma$ & $E_\gamma$\,(MeV) \\ \hline
$^{15}\text{N}^*$& 5.27 (5.30) & $5/2^+$ ($1/2^+$)& 0.09 & 1.00 & 5.27 \\
                 & 6.32 & $3/2^-$ & 0.87 & 1.00 & 6.32 \\
                 & 9.93 & $3/2^-$ & 0.04 & 0.02 & $2.63 + 7.30$ \\
                 &      &         &      & 0.04 & $3.61 + 6.32$ \\
                 &      &         &      & 0.13 & $4.63 + 5.30$ \\
                 &      &         &      & 0.13 & $4.66 + 5.27$ \\
                 &      &         &      & 0.65 & $9.93$ \\
$^{15}\text{O}^*$& 5.18 (5.24) & $1/2^+$ ($5/2^+$) & 0.09 & 1.00 & 5.18 \\
                 & 6.18 & $3/2^-$ & 0.91 & 1.00 & 6.18 \\
\hline \hline
\end{tabular*}
\end{table*}

Recently a new carbon SF with high missing energy resolution, capable of distinguishing low-lying discrete excited states, was published~\cite{PhysRevC.110.054612}.
The new SF directly determines the discrete excited state to which the nucleus transitions after interaction.
Therefore, manual assignment of discrete excited states within \textsc{NucDeEx} is no longer necessary.
This approach simplifies the enforcement of energy conservation and enables a more physically consistent description of the nuclear deexcitation process.
Accordingly, version 2.2 introduces a dedicated mode for use with this new carbon SF.
Instead of randomly selecting the level as in Fig.~\ref{fig:algorithm} and Table~\ref{tab:br_p_16O}, it transitions to the level nearest to the excitation energy given by the new SF, followed by gamma-ray emission.

\subsection{Interfaces}
\textsc{NucDeEx} offers interfaces for \textsc{Geant4} and \textsc{NEUT}.
\textsc{Geant4} provides its own deexcitation model, G4PreCompoundModel~\cite{G4PreCo}.
By default, this model is coupled with the INCL++~\cite{PhysRevC.87.014606,Leray_2013} and BIC (Binary cascade)~\cite{Folger2004} cascade models.
In contrast, the BERT (Bertini) cascade model uses a simpler deexcitation model~\cite{WRIGHT2015175}.
\textsc{NucDeEx} provides an interface to couple with INCL++ within the \textsc{Geant4} framework.
\textsc{Geant4} version 10.5.1 integrated with \textsc{NucDeEx} is publicly available on GitHub~\cite{code_g4}.
Note that this is not an official distribution of \textsc{Geant4}.
\par
The interface for \textsc{NEUT} is also developed aiming at combined use with the SF model~\cite{BENHAR1994493,PhysRevD.72.053005,PhysRevD.111.033006}.
The interface has been migrated into the \textsc{NEUT} source code and is no longer included in the \textsc{NucDeEx} distribution~\cite{code}.
Further details of these applications are described in Sec.~\ref{sec:applications}.

\section{Comparisons with other deexcitation event generators} \label{sec:validation}
The performance of deexcitation event generators, including \textsc{NucDeEx}, was evaluated through comparisons with experimental data.
This study focuses on their ability to reproduce energy spectra and branching ratios of decay particles, thus high-excitation energy region above 16\,MeV is investigated.
\par
Table~\ref{tab:generators} lists the event generators compared in this study.
These generators are based on different theoretical models.
The Hauser-Feshbach (HF) model~\cite{PhysRev.87.366} and the Weisskopf-Ewing (WE) model~\cite{PhysRev.57.472} are statistical models in which each decay emits exactly one particle.
If the excitation energy is sufficiently high, sequential decays occur.
The WE model does not consider angular momentum conservation, whereas the HF model takes it into account to address this issue~\cite{ADAMCZYK201321}.
Therefore, the HF model is generally considered to provide more accurate descriptions than the WE model in this aspect.
GEM is a simulation code based on the WE model~\cite{GEM}.
The Fermi breakup (FB) model~\cite{Fermi:1950jd,BONDORF1995133} follows a different prescription, where all decays happen simultaneously.
Because it calculates all possible decays beforehand, the FB model is generally used for light nuclei with few decay channels.
The G4PreCompoundModel combines several models: GEM, WE, and FB.
One of these models is selected depending on the emitted particle species, nucleus, and excitation energy.
The INCL++~\cite{PhysRevC.87.014606,Leray_2013} provides several built-in deexcitation event generators, such as the FB model, \textsc{ABLAv3p}~\cite{PhysRevC.105.014623,PhysRevC.98.021602,BENLLIURE1998458,JUNGHANS1998635}, \textsc{GEMINI++}~\cite{PhysRevC.82.014610}, and statistical multifragmentation model (\textsc{SMM})~\cite{BONDORF1995133}.
The FB model in \textsc{INCL++} is essentially derived from \textsc{Geant4}.
\textsc{GEMINI++} and \textsc{SMM} are excluded from comparisons because they do not simulate gamma-ray emissions, which are essential for various neutrino experiments. 
In addition to publicly available generators, results from TALYS-based calculations by S.~Abe {\it et al.}~\cite{PhysRevD.107.072006} and Hu {\it et al.}~\cite{HU2022137183}, as well as CASCADE-based calculations by Yosoi {\it et al.}~\cite{YOSOI2003255,Yosoi2004} are included for comparisons.

\begin{table*}[htbp] \centering
\caption{A list of generators compared with experimental data.
        }
\label{tab:generators}
\begin{tabular*}{1.0\textwidth}{@{\extracolsep{\fill}}lll} \hline \hline
Generators & Model & Features \\ \hline 
\textsc{NucDeEx} & HF &  TALYS-based. \\
\textsc{INCL++/FB} & FB & Default model for light nuclei ($A\leq16$) in INCL++~\cite{PhysRevC.87.014606,Leray_2013}. \\
\textsc{INCL++/ABLAv3p}~\cite{PhysRevC.105.014623,PhysRevC.98.021602,BENLLIURE1998458,JUNGHANS1998635} & WE & Alternative model in INCL++. \\ 
G4PreCompoundModel~\cite{G4PreCo} & GEM, WE, and FB & Geant4 custom deexcitation model. \\
\textsc{TALYS} (Abe {\it et al.})~\cite{PhysRevD.107.072006} & HF & Closed-source.\\
\textsc{TALYS} (Hu {\it et al.})~\cite{HU2022137183} & HF & Closed-source. \\
\textsc{CASCADE} (Yosoi {\it et al.})~\cite{YOSOI2003255,Yosoi2004} & HF & Closed-source.\\
\hline \hline
\end{tabular*}
\end{table*}

For simulations using \textsc{NucDeEx}, \textsc{INCL++/FB}, \textsc{INCL++/ABLAv3p}, and G4PreCompoundModel, the excitation energy $E_x$ was calculated by subtracting the proton separation energy $S_p$ from the removal energy $\tilde{E}$ determined by the previous SF~\cite{BENHAR1994493,PhysRevD.72.053005}
\begin{align}\label{eq:Ex}
 E_x = \tilde{E} - S_p
\end{align}
Since the new carbon SF~\cite{PhysRevC.110.054612} is essentially equivalent to the previous SF in the high-excitation energy region, the choice between them is not significant for the discussion in this section.
For other closed-source generators, the branching ratios reported in their papers were used.
Since the comparisons focus on excitation energies above 16\,MeV, the updates for discrete excited states described in Sec.\ref{sec:phole} do not affect the \textsc{NucDeEx} results. 
Other minor updates slightly change the results compared to those presented in Ref.\cite{PhysRevD.109.036009}.
\par
In addition to generators compared in this paper, \textsc{GEMINI++4$\nu$}, a tuned version of \textsc{GEMINI++}, was recently released~\cite{NIU2025139203}.
\textsc{GEMINI++4$\nu$}, mostly based on original GEMINI++, introduced suppression parameters for the particle decay widths and adjusted these parameters to reproduce experimental data.
The suppression parameters were defined individually for each particle and are independent of the excitation energy.
The experimental data~\cite{YOSOI2003255,Yosoi2004,YosM:2003} discussed in Sec.~\ref{sec:validation_data} were used for tuning, resulting in the best overall agreement among the generators used in this paper as intended.
This paper focuses on the benchmarking untuned generators, excluding \textsc{GEMINI++4$\nu$} from the comparisons.

\subsection{Energy spectra predicted by generators} \label{sec:Espe}
\begin{figure}[htbp] \centering
\includegraphics[width=1.0\columnwidth]{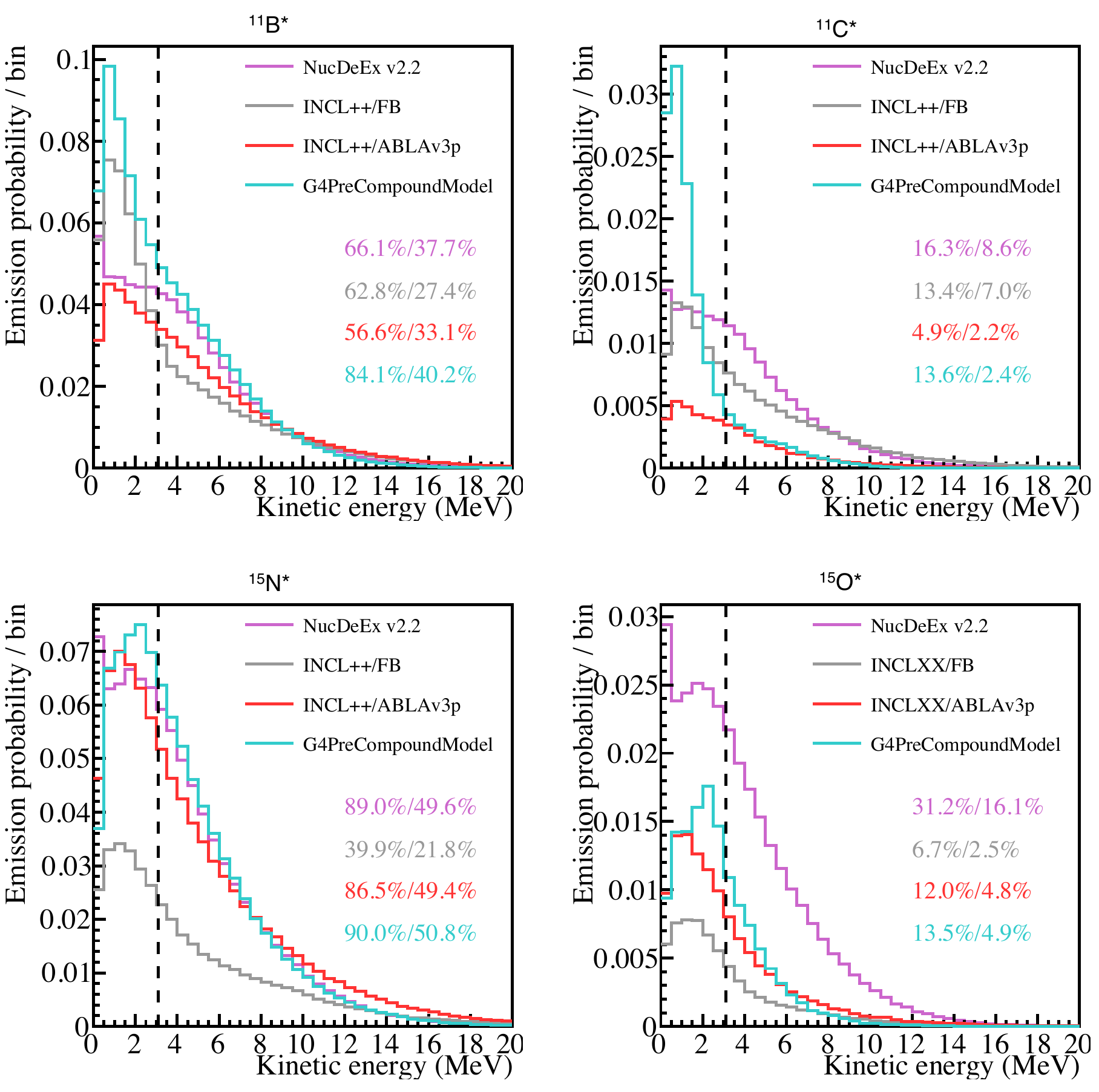}
\caption{Neutron energy spectra from deexcitation of $^{11}\text{B}^*$ (top left), $^{11}\text{C}^{*}$ (top right), $^{15}\text{N}^{*}$ (bottom left), and $^{15}\text{O}^{*}$ (bottom right) predicted by event generators.
         Excitation energy is selected to 16--35\,MeV for $^{11}\text{B}^*$ and $^{11}\text{C}^{*}$, 
         while it is 20--40\,MeV for $^{15}\text{N}^{*}$ and $^{15}\text{O}^{*}$.
         The vertical dashed lines represent the detection energy thresholds in the experiment at RCNP~\cite{YOSOI2003255,Yosoi2004,YosM:2003}.
         The numbers shown in each panel represent the emission probability of each event generator:
         total emission probability and that above the threshold are given in this order.
         All decays both of single-step and multistep are included.
        }
\label{fig:Espe_neutron}
\end{figure}

Particles emitted by the deexcitation process have relatively low energies of several MeV compared to those emitted by direct quasielastic (QE) interactions, which are typically around 100\,MeV.
For example, the typical excitation energies of $s_{1/2}$-hole states of $^{11}\text{B}^*$ and $^{15}\text{N}^*$ are approximately 20--40\,MeV.
In most cases, sequential decay occurs rather than the emission of a single high-energy particle.
Therefore, most of the excitation energy is consumed by the separation energy, which is typically about 10\,MeV, leaving the emitted particles with several MeV.
Figure~\ref{fig:Espe_neutron} shows neutron energy spectra predicted by deexcitation event generators.
Energy spectra for other particles are shown in the Appendix.
Although it depends on the particles species and generators, it is observed that 40\%--50\% of the decay particles have low energies below the detection threshold in the experiment with a proton beam at Osaka University’s Research Center for Nuclear Physics (RCNP)~\cite{YOSOI2003255,Yosoi2004,YosM:2003}.
These experiments, which are referred to as normal kinematics experiments, therefore have a limitation in detecting low-energy decay particles.
On the other hand, experiments using an ion beam, referred to as inverse kinematics experiments, such as GSI~\cite{PANIN2016204} and SAMURAI~\cite{Shimizu_2011}, can measure decay particles without detection thresholds because the center-of-mass frame is boosted.
Thus, inverse kinematics experiments are more effective for measuring the deexcitation process.
\par
Both the total emission probabilities and energy spectral shapes vary significantly depending on the generators used.
This indicates that when tuning the model, it is not sufficient to look only at the particles above the threshold measured at normal kinematics experiments.
And comprehensive measurements of the low-energy components using inverse kinematics with an ion beam are essential for proper model tuning.
Accurate tuning including energy spectra would require a method that consider excitation energy dependence, such as adjusting level density parameters, rather than simply adjusting suppression parameters for each particle.

\subsection{Comparisons with experimental data} \label{sec:validation_data}
\begin{figure}[htbp] \centering
\includegraphics[width=1.0\columnwidth]{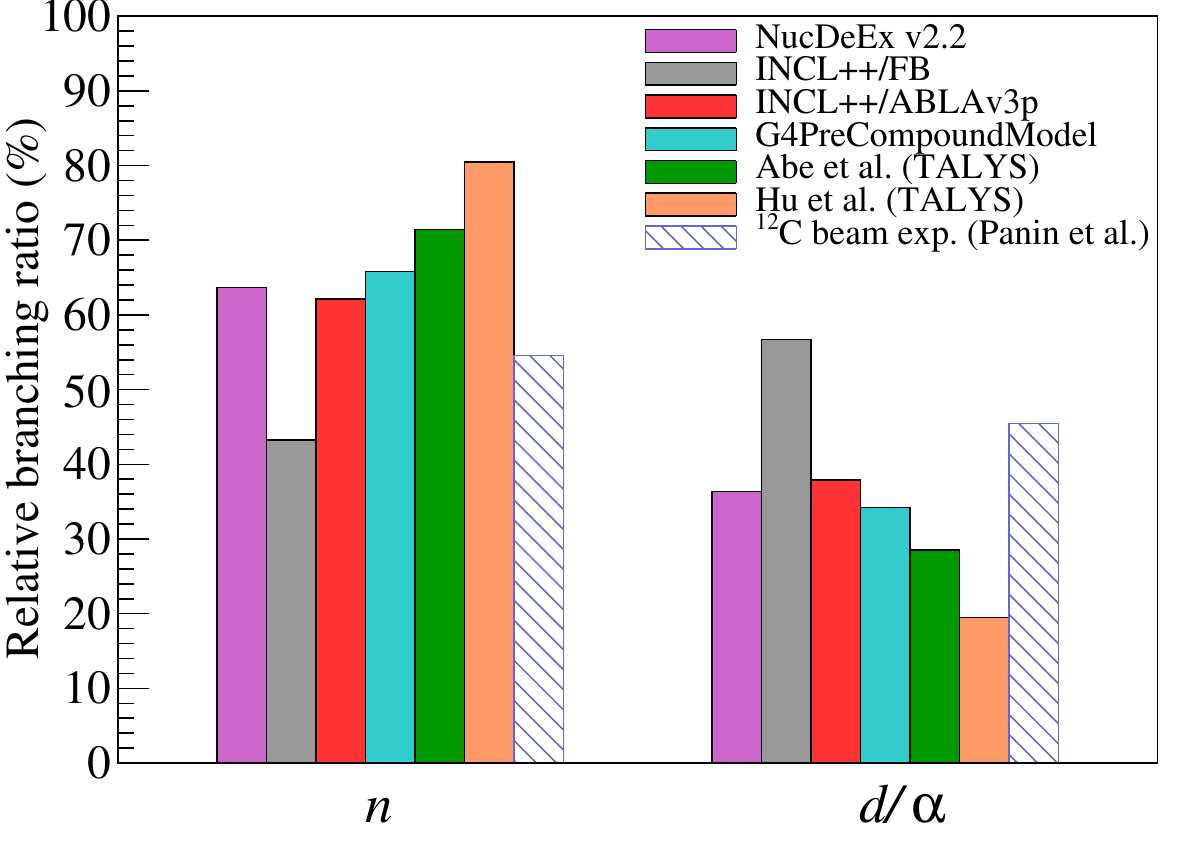}
\caption{Comparison of relative branching ratios of $n$ and $d/\alpha$ for $^{11}$B$^*$ with excitation energies of 16--35\,MeV.
         The magenta histograms show the results using \textsc{NucDeEx}.
         The green and orange histograms show the results from Ref.~\cite{PhysRevD.107.072006} and Ref.~\cite{HU2022137183} using \textsc{TALYS}, respectively.
         Experimental data~\cite{PANIN2016204} is shown as blue-hatched histogram.
         The branching ratios shown here account for only single-step decays.
        }
\label{fig:11B_nda}
\end{figure}

Figure~\ref{fig:11B_nda} shows a comparison of relative branching ratios of $n$ and $d/\alpha$ in $^{12}\text{C}(p,2p)^{11}\text{B}^*$ reactions measured at GSI~\cite{PANIN2016204}.
This experiment used $^{12}$C beam at $400$\,MeV/u.
Because the center-of-mass frame is boosted, even low-energy decay particles have sufficient energy in the laboratory frame to be detected without thresholds.
The FB model tends to overestimate $\alpha$ particle emission, resulting in an inconsistent trend with the experiment and other generators: the relative branching ratio of deuterons and $\alpha$ particles exceeds that of neutrons.

\begin{figure*}[htbp] \centering
\begin{minipage}[b]{0.49\textwidth} \centering
\includegraphics[width=1.0\columnwidth]{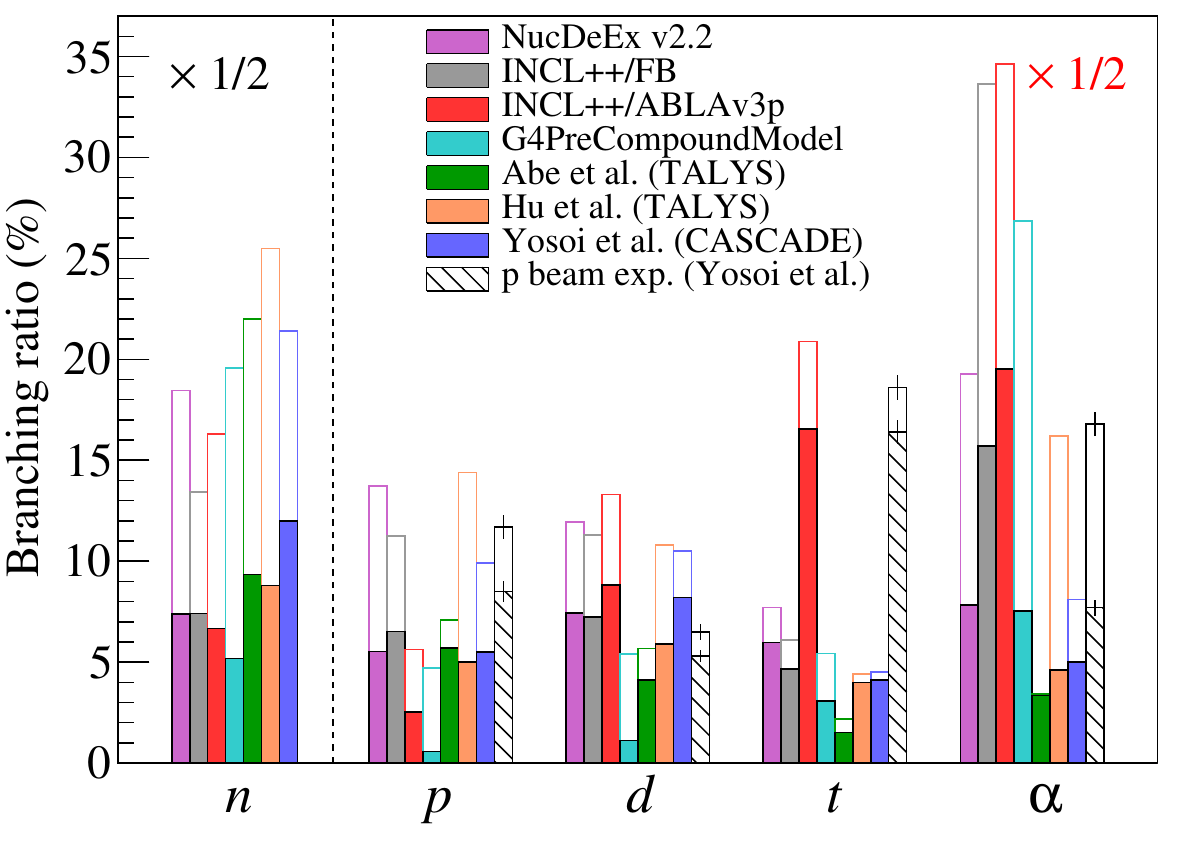}
\end{minipage}
\begin{minipage}[b]{0.49\textwidth} \centering
\includegraphics[width=1.0\columnwidth]{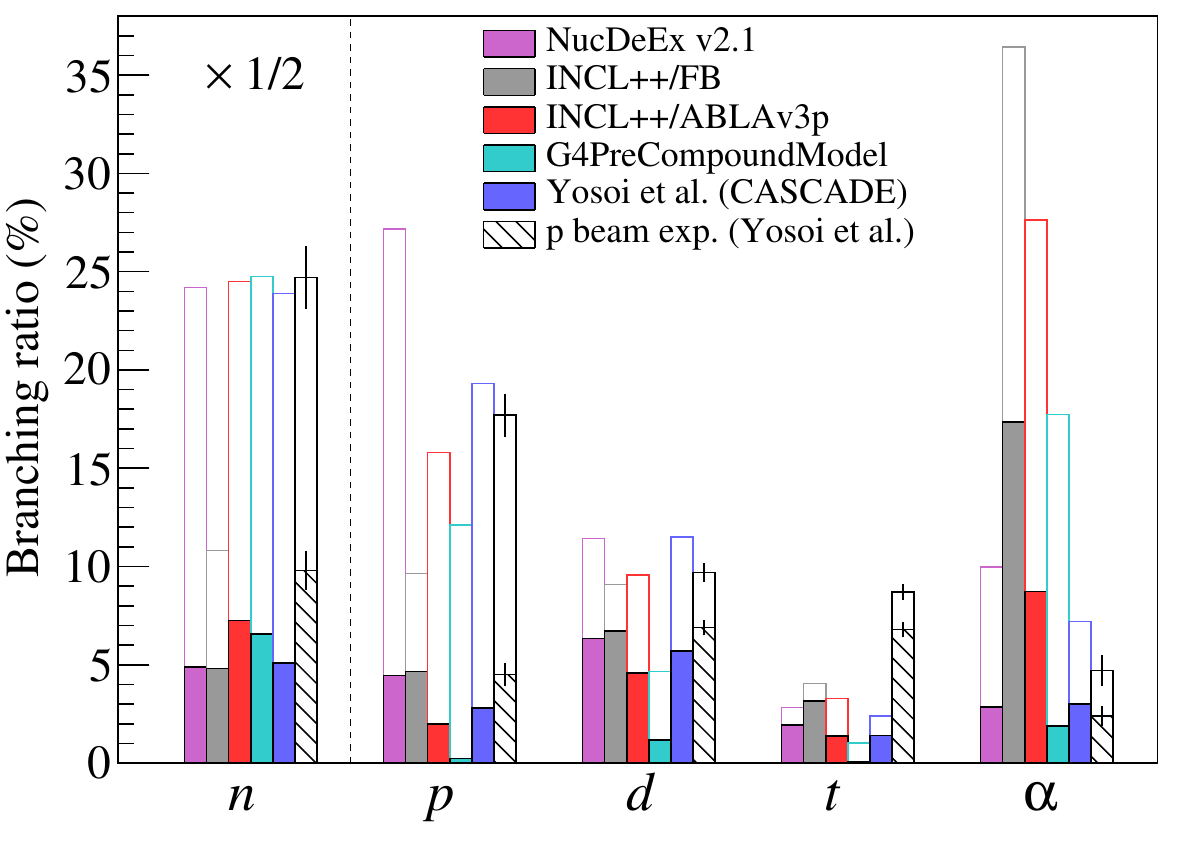}
\end{minipage}
\caption{Comparison of measured and predicted branching ratios of $n$, $p$, $d$, $t$, and $\alpha$ for $^{11}$B$^*$ with 16--35\,MeV excitation energy (left) and for $^{15}$N$^*$ with 20--40\,MeV excitation energy (right).
         The branching ratios of $n$ are scaled by a factor of 1/2.
         The branching ratio of \textsc{ABLA}'s $\alpha$ from $^{11}\text{B}^*$ is also scaed by a factor of 1/2.
         The magenta histograms show the results using NucDeEx.
         The green and orange histograms show the results from Ref.~\cite{PhysRevD.107.072006} and Ref.~\cite{HU2022137183} using \textsc{TALYS}, respectively.
         The black histograms denote experimental data from Yosoi {\it et al.} with statistical errors,
         and the authors provide the prediction using CASCADE code written with the blue histograms~\cite{YOSOI2003255,Yosoi2004}.
         The hatched or filled histograms represent the branching ratios for single-step decays,
         and the open histograms represent those for multistep decays.
         }
\label{fig:11B_15N_br}
\end{figure*}

Figure~\ref{fig:11B_15N_br} shows comparisons of absolute branching ratios measured in $^{12}\text{C}(p,2p)^{11}\text{B}^*$ and $^{16}\text{O}(p,2p)^{15}\text{N}^*$ reactions at RCNP~\cite{YOSOI2003255,Yosoi2004,YosM:2003}.
This experiment used a proton beam of 392\,MeV.
In this experiment, detection thresholds of 3.1--4.6\,MeV were applied to the decay particle measurement, resulting in 40\%--50\% inefficiency, as presented in Sec.~\ref{sec:Espe} and the Appendix.
The branching ratio for decay channel $i$, $Br_i$, is given by
\begin{align} \label{eq:br_yosoi}
Br_i = \frac{\int n_i (4\pi/\Delta\Omega) dE_x}{\int N dE_x}
\end{align}
where $E_x$ is the excitation energy, $N$ is the number of $(p,2p)$ events, $\Delta\Omega$ is the total solid angle of the SSD detector for measuring decay particles, and $n_i$ is the number of decay particles detected for channel $i$.
Since $n_i$ includes only detected particles, the branching ratios represent only the components above the detection thresholds.
Note that the acceptance of the SSD detector was limited to $\Delta\Omega = 3.5\%$ of $4\pi$, and thus, sequential decays are practically undetectable.
As a result, the branching ratio $Br_i$ can exceed 100\%.
In the analysis, invariant mass was reconstructed by summing the excitation energy measured by $(p,2p)$ kinematics and energy of decay particle measured by SSD.
If events have smaller invariant mass than the true target mass, it indicates the presence of undetected decay particles.
This method allows one to distinguish between single-step and multistep decays without detecting multiple decay particles directly (See Refs.~\cite{YOSOI2003255,Yosoi2004,YosM:2003} for details).
The same analysis is applied to simulation to ensure fair comparisons with data.
\par
To systematically evaluate the performance of these generators, chi-squared values are calculated for each generator (Table~\ref{tab:chi2}).
The chi-squared is defined as 
\begin{align}
\chi^2 = \sum_i \left( \frac{Br_i^{MC}-Br_i^{Data}}{\Delta Br_i^{Data}} \right)^2
\end{align}
where $\Delta Br_i^{Data}$ is the statistical uncertainty of the branching ratios for experimental data, $Br_i^{MC}$ and $Br_i^{Data}$ are the branching ratios for decay channel $i$ of MC and experimental data, respectively.
Both single step and multistep decays are taken into account in this calculation.
Note that only statistical uncertainty is considered since the systematic uncertainty of the experiment is not provided.
Thus, only the relative values of chi-squared are meaningful for evaluating the performance of the generators.
It is found that \textsc{INCL++/ABLA} exhibits an energy conservation problem in the decay channel of $^{11}\text{B}^*\rightarrow t +\alpha+\alpha$, leading to a sizable overestimation of the branching ratio for this channel.
Although the branching ratio of tritons appears to be close to the experimental value, this agreement is attributed to the violation of energy conservation.
Therefore, this is not considered to be indicative of physical accuracy.
\textsc{INCL++/FB} overestimates the branching ratio for $\alpha$ particles, while underestimating neutron emissons.
G4PreCompoundModel also overestimates the branching ratio for $\alpha$, mainly due to the FB model selected in low-excitation energy region.
As observed in Fig.~\ref{fig:11B_nda}, the FB model consistently overpredicts $\alpha$ emissions, which does not agree well with experimental data.
Furthermore, G4PreCompoundModel underpredicts single-step proton decays, which also fails to reproduce the data.
As a result, \textsc{NucDeEx} shows the best, or at least comparable, agreement with experiments in terms of $\chi^2$.
Generators based on the HF model tend to reproduce the experimental data better than others.
This is consistent with expectations since the HF model includes angular momentum conservation.

\begin{table*}[htbp] \centering
\caption{The chi-squared values of each generators to the data shown in Fig.~\ref{fig:11B_15N_br}.
         Since the systematic uncertainty is not provided, the statistical uncertainty given by the experiment is only considered.
         }
\label{tab:chi2}
\begin{tabular*}{0.7\textwidth}{@{\extracolsep{\fill}}llll} \hline \hline
 & & \multicolumn{2}{c}{$\chi^2/ndf$} \\
Generator      &  Model & $^{11}\text{B}^*$ & $^{15}\text{N}^* $ \\ \hline
\textsc{NucDeEx} v2.2 & HF & 483/8 & 279/10 \\
\textsc{INCL++/FB} & FB & 1028/8 & 1387/10 \\
\textsc{INCL++/ABLAv3p} & WE & 7322/8 & 723/10 \\
G4PreCompoundModel & GEM, WE, and FB & 1181/8 & 767/10 \\
\textsc{TALYS} (Abe {\it et al.}) & HF & 947/8 & - \\
\textsc{TALYS} (Hu {\it et al.})  & HF & 674/8 & - \\
\textsc{CASCADE} (Yosoi {\it et al.}) & HF & 676/8 & 263/10 \\
 \hline \hline
\end{tabular*}
\end{table*}

Table~\ref{tab:15N_gamma} shows a comparison of gamma-ray branching ratios from $^{15}\text{N}^*$ with experimental data measured at RCNP~\cite{kobayashi2006deexcitation}.
This experiment measured gamma rays of $^{16}\text{O}(p,2p\gamma)$ reaction using 392\,MeV proton beam.
A clear discrepancy is observed in \textsc{INCL++/ABLA}.
Since it does not properly consider low-lying discrete excited states, it emits gamma rays with energies above 7.4 MeV (see Appendix).
This issue is also pointed out in Ref.~\cite{hino2025simulationmodelinvestigationneutronoxygen}.
The result of \textsc{INCL++/ABLA}, showing significant discrepancies with the experimental data, indicates that it is not suitable for low-energy analyses such as DSNB searches. 
The \textsc{INCL++/FB} reproduces the experiment most accurately. 
However, as shown previously, it gives significant discrepancy in neutron and $\alpha$ branching ratios.
Therefore, it is considered unsuitable for DSNB searches, which require accurate prediction of neutron multiplicity.
\textsc{NucDeEx} achieves comparable or superior agreement with gamma-ray branching ratios relative to G4PreCompoundModel.
Overall, \textsc{NucDeEx} shows good reproducibility across all comparisons discussed here.
This infers that using \textsc{NucDeEx} instead of G4PreCompoundModel in Geant4 will enable more accurate description of deexcitation.
The application of \textsc{NucDeEx} to \textsc{Geant4} will be described in Sec.~\ref{sec:geant4}.

\begin{table*}[htbp] \centering
\caption{Comparison of gamma-ray branching ratios with total energy of 3\,MeV$ < E_{\gamma,\text{tot}} < 6$\,MeV and 6\,MeV $< E_{\gamma,\text{tot}}<7.4$\,MeV.
         The experimental data was measured using proton beams at RCNP~\cite{kobayashi2006deexcitation}.
         Excitation energies of 16--40\,MeV are selected in this comparison.
         }
\label{tab:15N_gamma}
\begin{tabular*}{0.65\textwidth}{@{\extracolsep{\fill}}lll} \hline \hline
 & \multicolumn{2}{c}{Banching ratio (\%)} \\
 & 3\,MeV$ < E_{\gamma,\text{tot}} < 6$\,MeV & 6\,MeV$< E_{\gamma,\text{tot}} < 7.4$ MeV \\ \hline
\textsc{NucDeEx} v2.2 & 31.1 & 8.5 \\
\textsc{INCL++/FB} & 31.1 & 16.4 \\
\textsc{INCL++/ABLAv3p} & 0 & 0 \\
G4PreCompoundModel & 22.9 & 8.7 \\
Experiment & $27.9\pm1.5^{+3.4}_{-2.6}$ & $15.6\pm1.3^{+0.6}_{-1.0}$ \\ \hline\hline
\end{tabular*}
\end{table*}

\section{Applications} \label{sec:applications}
\textsc{NucDeEx} is a general-purpose deexcitation event generator that can be integrated into other simulations, such as \textsc{Geant4} and \textsc{NEUT}.
This section presents its applications and results.

\subsection{Geant4}\label{sec:geant4}
\textsc{Geant4} provides its own custom deexcitation model, G4PreCompoundModel~\cite{G4PreCo}.
By default, it is coupled with INCL++ and BIC cascade models, while the BERT uses its own simplified model.
As shown in Sec.~\ref{sec:validation}, \textsc{NucDeEx} shows better overall agreement with the data than G4PreCompoundModel.
Thus, incorporating \textsc{NucDeEx} into \textsc{Geant4} is expected to improve the accuracy of simulations.
A modified version of \textsc{Geant4} 10.5.1 incorporating \textsc{NucDeEx} is publicly available on GitHub~\cite{code_g4}.
This is not an official distribution of \textsc{Geant4}.
\par
\begin{figure*}[htbp] \centering
\begin{minipage}[b]{0.45\textwidth} \centering
\includegraphics[width=1.0\textwidth]{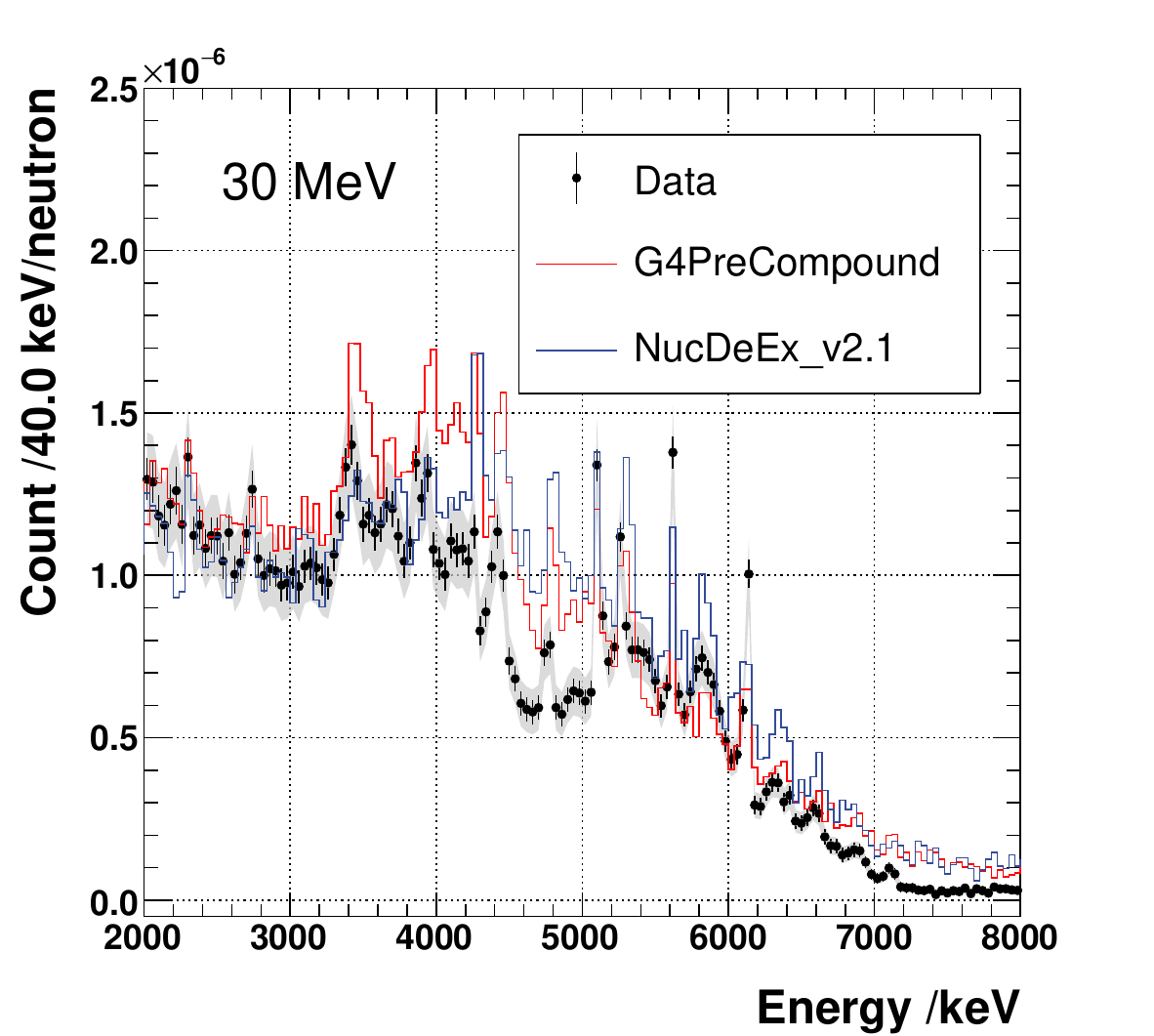}
\end{minipage}
\begin{minipage}[b]{0.45\textwidth} \centering
\includegraphics[width=1.0\textwidth]{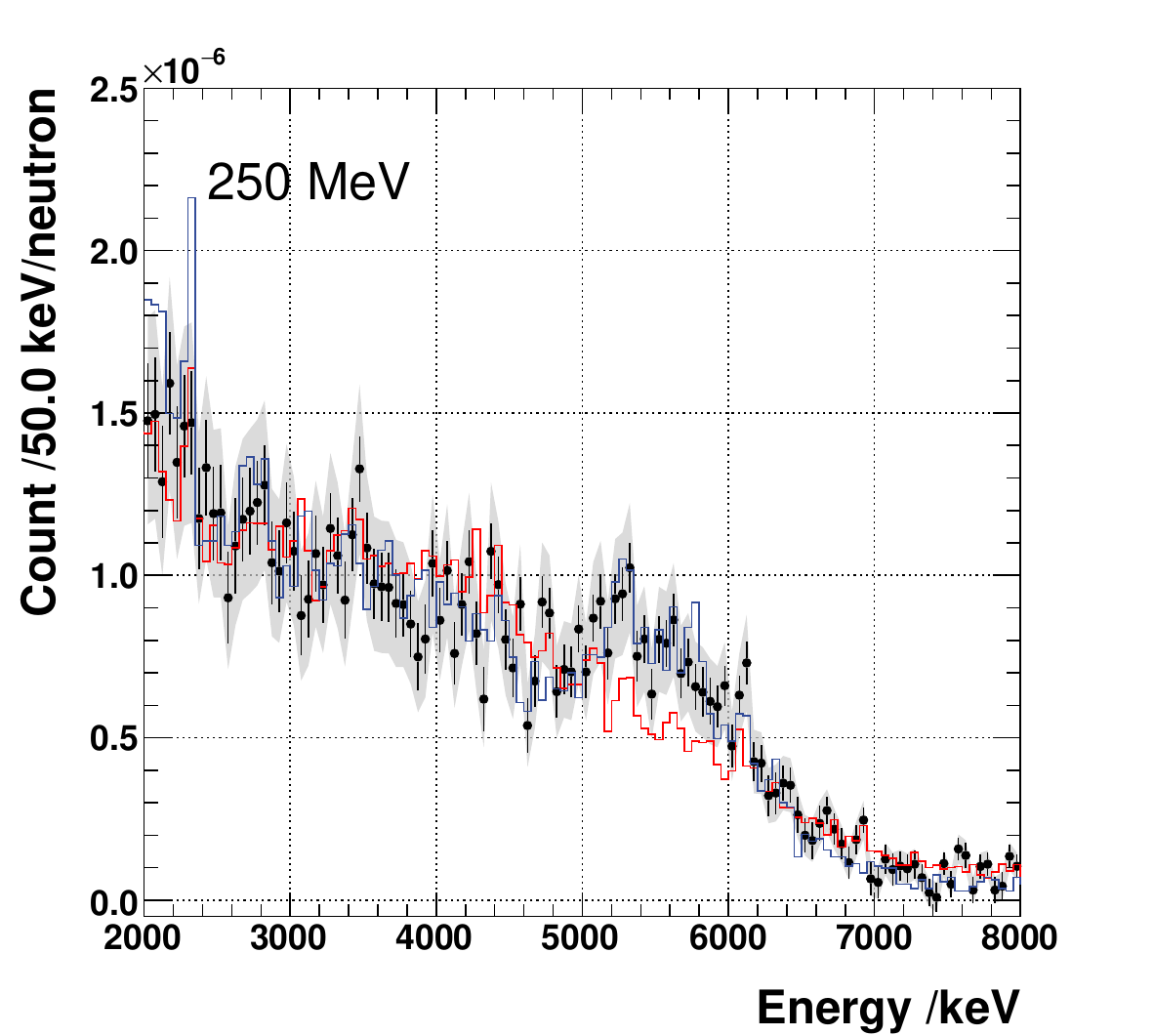}
\end{minipage}
\caption{The observed and predicted gamma-ray energy spectra from neutron-oxygen interaction at 30\,MeV and 250\,MeV of neutron beam energies.
The black dots represent the experimental data of E525~\cite{10.1093/ptep/ptae159}.
The red and dark blue lines are \textsc{Geant4} simulations using INCL++ coupled with G4PreCompoundModel and \textsc{NucDeEx}, respectively.
\textsc{NucDeEx} shows better agreement than G4PreCompoundModel in all neutron energies.
The figures are from Ref.~\cite{hino2025simulationmodelinvestigationneutronoxygen}, and the chi-squared values are summarized in the paper.
}
\label{fig:E525}
\end{figure*}
\begin{figure}[htbp] \centering
\includegraphics[width=0.9\columnwidth]{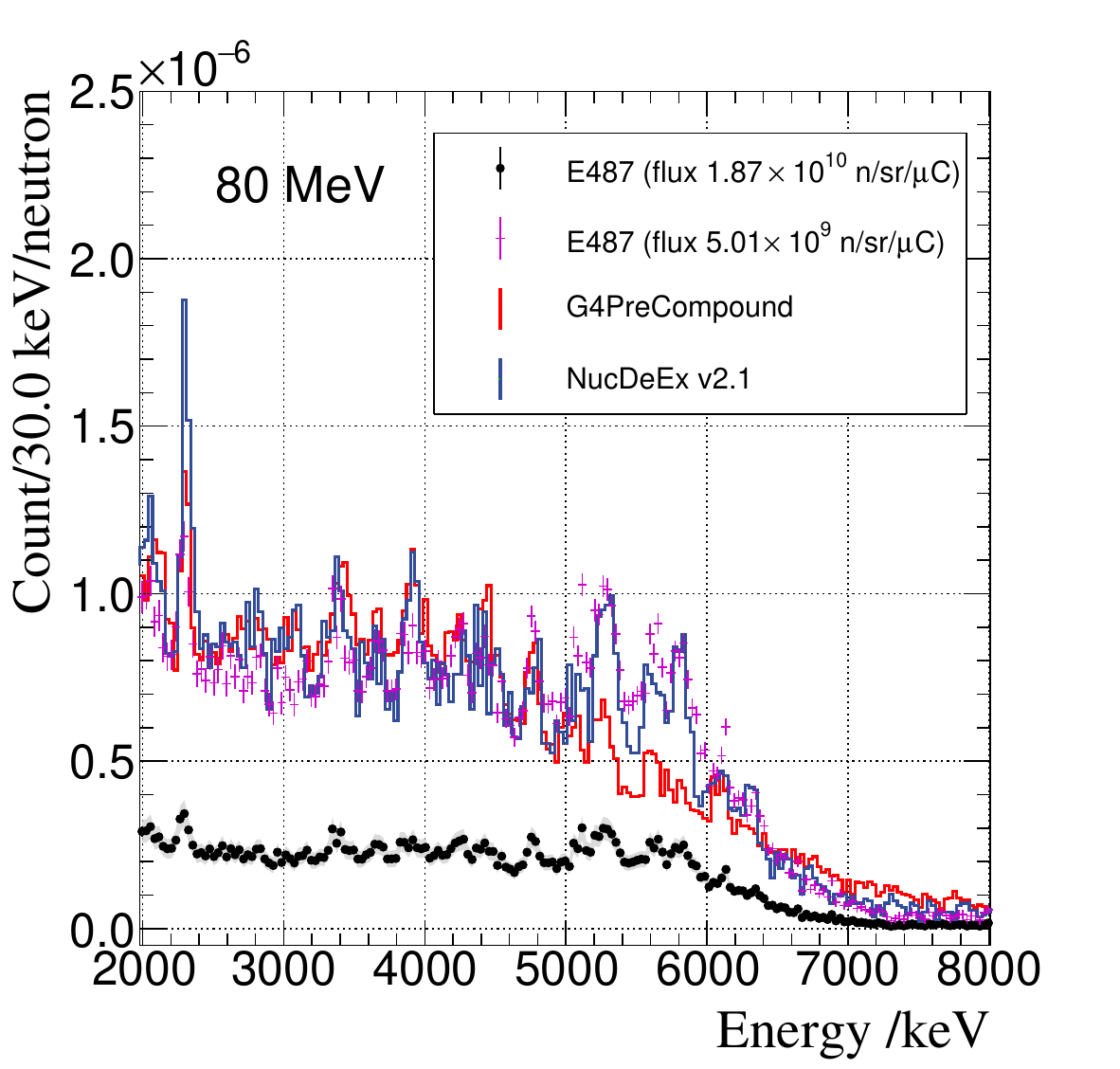}
\caption{The observed and predicted gamma-ray energy spectra from neutron-oxygen interaction at 80\,MeV neutron beam energies.
The black dots represent the experimental data of E487~\cite{PhysRevC.109.014620}, while Ref.~\cite{10.1093/ptep/ptae159} pointed out a potential issue in the flux normalization.
The magenta dots, representing the E487 data normalized with the E525 neutron flux at 250\,MeV, are also shown to avoid this issue.
The red and dark blue lines are \textsc{Geant4} simulations using INCL++ coupled with G4PreCompoundModel and \textsc{NucDeEx}, respectively.
The figures are from Ref.~\cite{hino2025simulationmodelinvestigationneutronoxygen}, and the chi-squared values are summarized in the paper.
}
\label{fig:E487}
\end{figure}

E487 and E525 experiments~\cite{PhysRevC.109.014620,10.1093/ptep/ptae159} measured gamma-ray energy spectra of $^{16}\text{O}(n,\gamma)$ inclusive reactions.
These experiments were conducted at RCNP using neutron beams with energies of 30\,MeV, 80\,MeV, and 250\,MeV on water target.
The observed gamma-ray energy spectra were compared with various combinations of cascade and deexcitation models~\cite{hino2025simulationmodelinvestigationneutronoxygen}.
Among the standard cascade models provided by Geant4 (BERT, BIC, INCL++), INCL++ shows the best agreement with the experimental data.
Replacing the default deexcitation model (G4PreCompoundModel) in INCL++ with NucDeEx led to further improvement in reproducibility across all neutron energies (Figs.~\ref{fig:E525} and \ref{fig:E487}).
Note that Ref.~\cite{10.1093/ptep/ptae159} pointed out a potential issue in the measurement of proton beam current in E487.
This may have resulted in an incorrect normalization of the neutron beam flux, as shown in Fig.~\ref{fig:E487}.
In particular, a significant improvement is observed at a neutron energy of 80\,MeV.
This energy region is important for detector simulations of atmospheric neutrino interactions, where the typical neutron energy is around 100\,MeV~\cite{Sakai:2024azr}.
NCQE interactions, which do not produce a charged lepton in the final state, can be a background in DSNB searches when the total energy of emitted gamma rays reaches approximately 20\,MeV.
This improvement at 80\,MeV is expected to directly enhance background estimations in DSNB searches at SK, especially for those from NCQE interactions.

\subsection{NEUT}\label{sec:neut}
\textsc{NucDeEx} has been officially integrated into the neutrino event generator \textsc{NEUT}~\cite{Hayato2021} and is available from version 5.9.0.
Currently, it is limited to QE interactions and is intended for used with the SF model~\cite{PhysRevD.111.033006}.
The \textsc{NEUT}-\textsc{NucDeEx} interface supports previous SF~\cite{BENHAR1994493,PhysRevD.72.053005} as well as the new carbon SF~\cite{PhysRevC.110.054612}, which has recently been implemented into \textsc{NEUT}.
The excitation energy is determined according to Eq~(\ref{eq:Ex}), neglecting the effects of the final-state cascade interactions.
In contrast, nuclear species are determined after the cascade by counting the outgoing nucleons.
Furthermore, a relativistic distorted wave impulse approximation (RDWIA) model has recently been implemented into \textsc{NEUT}~\cite{f7x5-snmz}
In this model, the missing energy distribution for each shell is parametrized using a Gaussian function, closely resembling that of the SF.
Therefore, \textsc{NucDeEx} is expected to be compatible with the RDWIA model as well.
\par
\textsc{NEUT} previously included a custom data-driven model based on measurements of $^{16}\text{O}(p,2p\gamma)$ at RCNP~\cite{kobayashi2006deexcitation}.
However, this data-driven model had several limitations.
First, the model is limited to $^{16}\text{O}$, and therefore it could not be used in scintillator detectors.
Second, due to the lack of branching ratio measurement for $^{15}\text{O}^*$, isospin symmetry was assumed.
This assumes that the neutron branching ratios of $^{15}\text{O}^*$ are identical to the proton branching ratios of $^{15}\text{N}^*$.
However, Ref.~\cite{PhysRevD.109.036009} showed that the isospin symmetry is violated due to the Coulomb potential.
Finally, the hole state was selected independently of the removal energy determined by the primary interaction, making the primary interaction and deexcitation entirely decoupled.
\textsc{NucDeEx} in \textsc{NEUT} is expected to resolve these limitations of the data-driven model.
It supports both $^{16}\text{O}$ and $^{12}\text{C}$ targets, enabling applications to water Cherenkov and scintillator detectors.
As a theory-based generator, it does not assume isospin symmetry and the Coulomb potential effects are incorporated through the HF model.
Furthermore, the excitation energy is derived from the removal energy given by the primary interaction, thereby linking the primary interaction and deexcitation process.
These features of the \textsc{NEUT}-\textsc{NucDeEx} interface are expected to allow more precise predictions of nuclear deexcitation associated with neutrino-nucleus interactions.
\par
For example, it is expected to improve background estimation in DSNB searches.
Reference~\cite{PhysRevD.103.053001, PhysRevD.103.053002,Cheng2025} discusses generator-dependent variations by combining a custom deexcitation model with existing neutrino event generators, \textsc{NuWro} and \textsc{GENIE}~\cite{ANDREOPOULOS201087}.
\textsc{NEUT} allows us to perform similar investigations using \textsc{NucDeEx} in its package.
Another application would be in the context of missing energy measurement using neutrinos from kaon decay-at-rest reported by JSNS$^2$ Collaboration~\cite{PhysRevLett.134.081801}.
It would be valuable to investigate how the predictions vary when \textsc{NucDeEx} is combined with the new carbon SF that has high missing energy resolution, which offers enforced connection of primary interaction and deexcitation.

\section{Conclusion and prospects} \label{sec:conclusion}
This paper presents updates on the nuclear deexcitation event generator \textsc{NucDeEx}, benchmarks against existing generators and experimental data, and applications to other simulations.
The treatment of low-lying discrete excited states was modified to mitigate bias arising from the determination of hole states and to reproduce the excited states observed in the experiments.
\textsc{NucDeEx} version 2.2 implements a specific mode designed for use in combination with a new carbon SF~\cite{PhysRevC.110.054612} that distinguishes low-lying discrete excited states.
These modifications enable a more accurate and physically consistent description of gamma-ray emission.
\par
Comparisons with other deexcitation event generators reveal significant variations in emission probabilities and energy spectra.
These discrepancies highlight the need for inverse kinematics experiments that can measure decay particles without practical detection thresholds, such as GSI~\cite{PANIN2016204} and SAMURAI~\cite{Shimizu_2011}.
Comparisons with experimental data show that \textsc{NucDeEx} gives the best or comparable agreement with experiments overall.
Note that \textsc{GEMINI++4$\nu$}~\cite{NIU2025139203}, which was tuned to match the experimental data from normal kinematics, shows better agreement than the generators used in this paper.
\par
\textsc{NucDeEx} can be used with \textsc{Geant4} and \textsc{NEUT}.
Simulation studies by E487 and E525 experiments using \textsc{Geant4}, which measured $^{16}\text{O}(n,\gamma)$ inclusive reactions, show improvements in the gamma-ray energy spectra by replacing the default deexcitation model G4PreCompoundModel with \textsc{NucDeEx}.
The most notable improvement was observed at a neutron energy of 80\,MeV, corresponding to the typical nucleon energy of atmospheric neutrinos.
Thus, \textsc{NucDeEx} is expected to improve background estimation for DSNB searches, where the atmospheric neutrinos are the dominant background.
\par
This paper focuses on generator benchmarking and does not address model tuning.
Fundamental tuning of the deexcitation model requires the following considerations.
First, inverse kinematics experiments should be conducted to measure decay particles without practical detection thresholds.
Given that generator dependence is highly evident in decay particle energy spectra shapes, low-energy components should be measured and taken into account.
Second, it may be preferable to tune the level density parameter to account for excitation energy dependence rather than simply introducing a single parameter like the suppression parameter.
Such advanced tuning cannot be achieved using published experimental data alone, and requires new experiments to collect sufficient information.

\begin{acknowledgments}
This work was supported by JSPS KAKENHI Grant No.~23KJ0319 and 25H00631.
The author expresses deep gratitude to the host researcher of this JSPS fellowship, Professor Yoshinari Hayato, for his invaluable assistance in improving the manuscript.
The author is grateful to the SAMURAI-79 Collaboration for valuable discussions on the deexcitation process and its simulations.
The author also thanks to Dr. Yota Hino for the use of \textsc{NucDeEx} in the simulation.
\end{acknowledgments}

\appendix*
\section{ENERGY SPECTRA OF DECAY PARTICLES} \label{sec:appen_Espe}
\begin{figure*}[htbp] \centering
\includegraphics[width=1.0\textwidth]{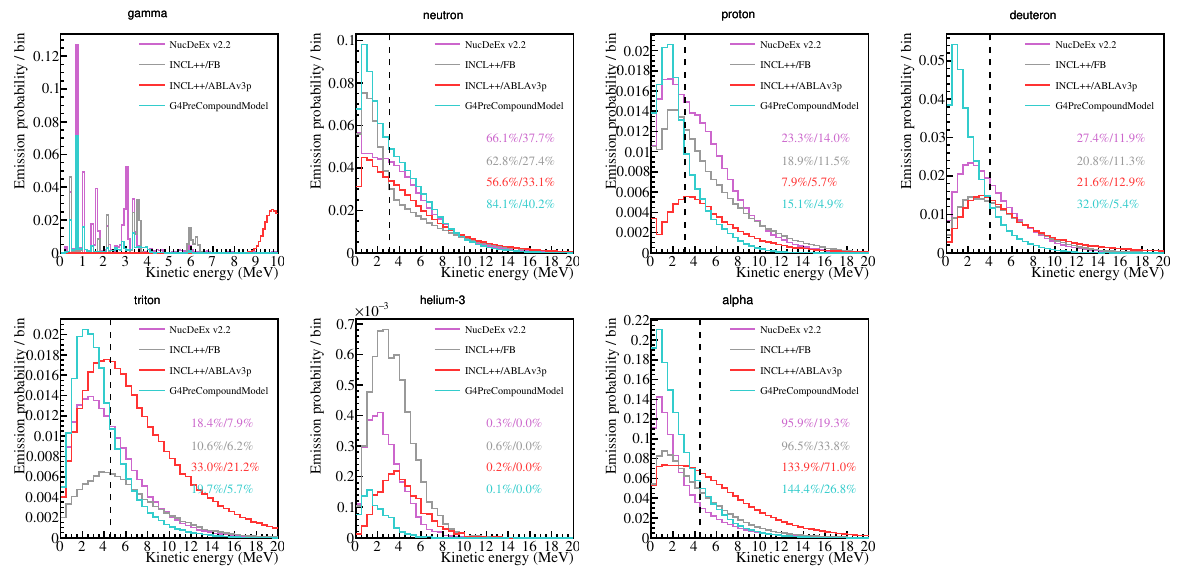}
\caption{Energy spectra of decay particles from $^{11}\text{B}^{*}$ with excitation energy of 16--35\,MeV predicted by deexcitation event generators.
         The vertical dashed lines represent the detection energy thresholds in the experiment at RCNP~\cite{YOSOI2003255,Yosoi2004,YosM:2003}.
         The numbers shown in each panel represent the emission probability of each particle:
         total emission probability and that above the threshold are given in this order for each event generator.
         All decays both of single-step and multistep are included.
         This emission probability can exceed 100\% when multiple particles are emitted.
        }
\label{fig:Espe_11B}
\end{figure*}

\begin{figure*}[htbp] \centering
\includegraphics[width=1.0\textwidth]{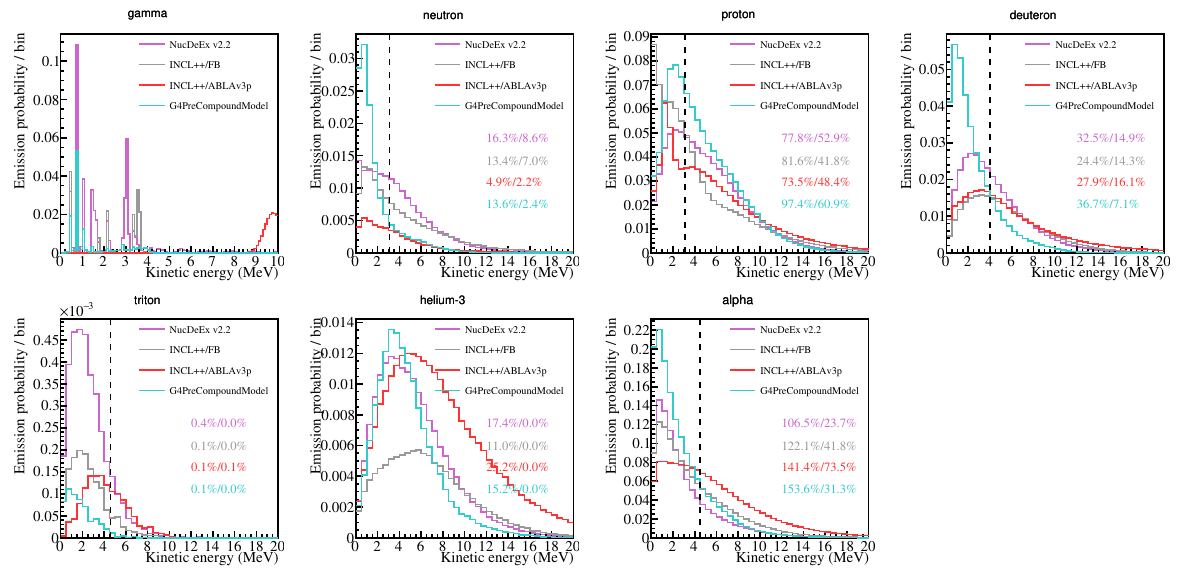}
\caption{The same as Fig.~\ref{fig:Espe_11B}, but from $^{11}\text{C}^{*}$ with excitation energy of 16--35\,MeV.}
\label{fig:Espe_11C}
\end{figure*}

\begin{figure*}[htbp] \centering
\includegraphics[width=1.0\textwidth]{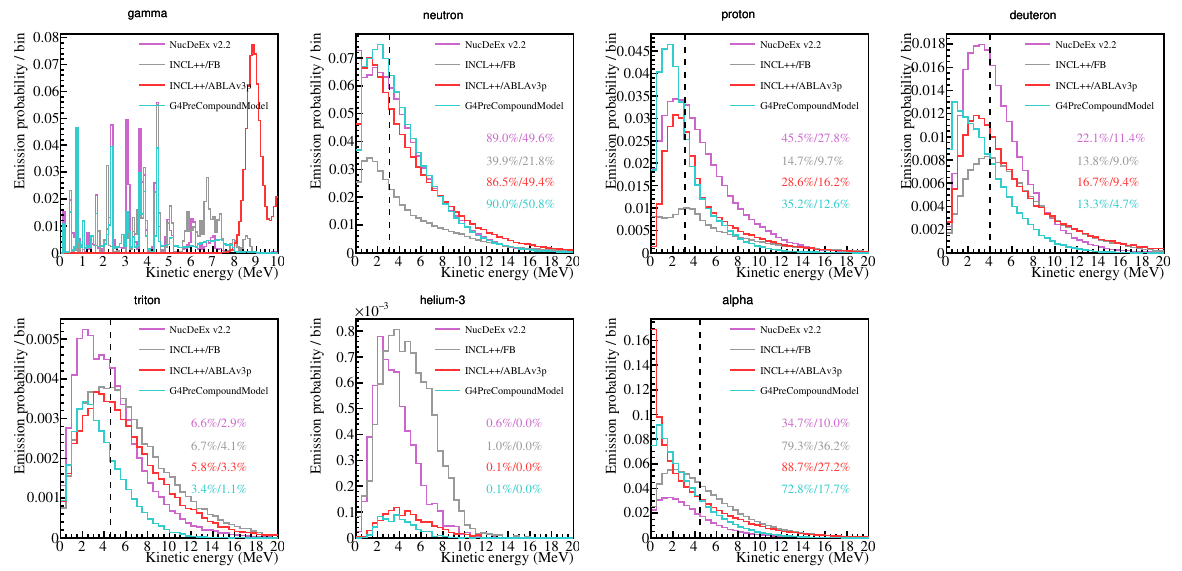}
\caption{The same as Fig.~\ref{fig:Espe_11B}, but from $^{15}\text{N}^{*}$ with excitation energy of 20--40\,MeV.}
\label{fig:Espe_15N}
\end{figure*}

\begin{figure*}[htbp] \centering
\includegraphics[width=1.0\textwidth]{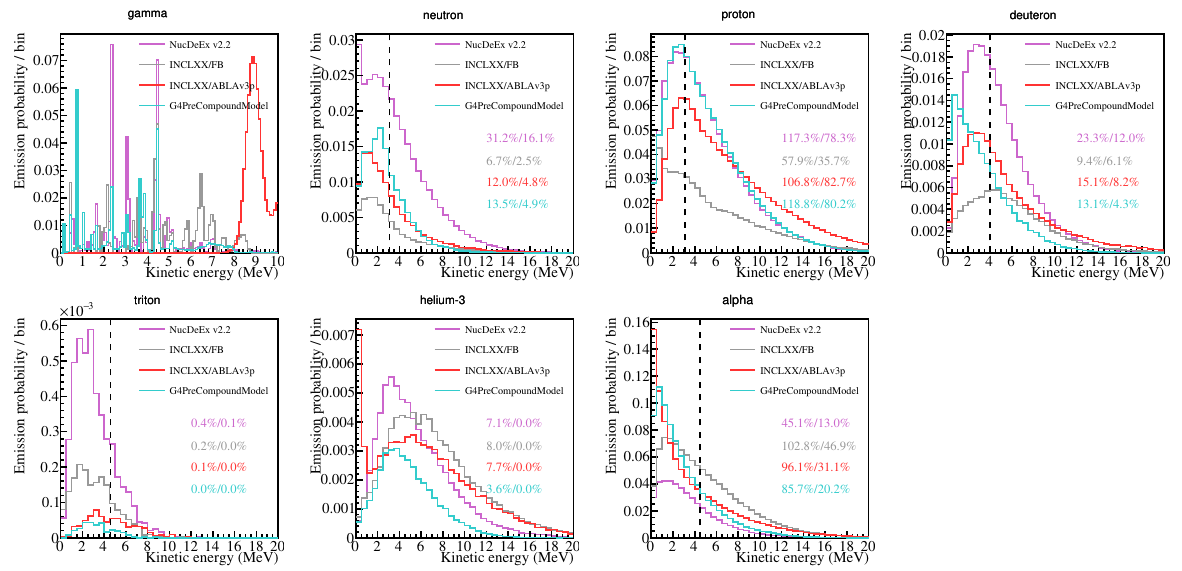}
\label{fig:Espe}
\caption{The same as Fig.~\ref{fig:Espe_11B}, but from $^{15}\text{O}^{*}$ with excitation energy of 20--40\,MeV.}
\label{fig:Espe_15O}
\end{figure*}

Figures~\ref{fig:Espe_11B}--\ref{fig:Espe_15O} shows predicted energy spectra of decay particles from $^{11}\text{B}^*$, $^{11}\text{C}^{*}$, $^{15}\text{N}^{*}$, and $^{15}\text{O}^{*}$.
The same excitation energy as Sec.~\ref{sec:validation} is assumed.
Although it depends on the particles and generators, it is observed that 40\%--50\% of decay particles have energies below the threshold.
Furthermore, not only the absolute total emission probability, but also the shape of the spectra vary largely depending on the generators.
The gamma-ray energy spectra predicted by \textsc{INCL++/ABLA} is above ~8\,MeV, indicating that low-lying discrete excited states are not taken into account.

\FloatBarrier
\bibliography{bib}

\end{document}